\documentclass[journal=jacsat,manuscript=article]{achemso}

\usepackage{subcaption}
\usepackage{float}
\usepackage{graphicx}
\usepackage{amsmath}
\author{Joel Creutzberg}
\affiliation[KULeuven]
{Department of Chemistry, KU Leuven, Celestijnenlaan 200F, B-3001 Leuven, Belgium}
\email{joel.creutzberg@kuleuven.be}

\author{Wojciech Skomorowski}
\affiliation[Warsaw]
{Centre of New Technologies, University of Warsaw, Banacha 2c, 02-097 Warsaw, Poland}

\author{Thomas-C. Jagau}
\affiliation[KU Leuven]
{Department of Chemistry, KU Leuven, Celestijnenlaan 200F, B-3001 Leuven, Belgium}
\title[An \textsf{achemso} demo]
{Computing decay widths of autoionizing Rydberg states 
with complex-variable coupled cluster theory}

\begin{document}

\begin{tocentry}
\includegraphics[width=0.7\textwidth]{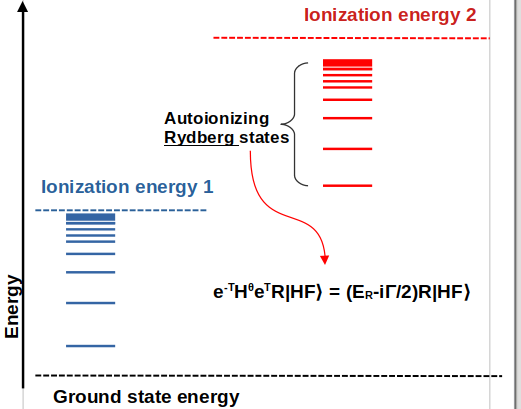}
\end{tocentry}

\begin{abstract}
We compute autoionization widths of various Rydberg states of neon 
and dinitrogen by equation-of-motion coupled-cluster theory combined 
with complex scaling and complex basis functions. This represents the 
first time that complex-variable methods are applied to Rydberg states 
represented in Gaussian basis sets. A new computational protocol based 
on Kaufmann basis functions is designed to make these methods applicable 
to atomic and molecular Rydberg states. As a first step, we apply our 
protocol to the neon atom and computed widths of the $3s$, $3p$, $4p$ 
and $3d$ Rydberg states. We then proceed to compute the widths of the 
$3s\sigma_g$, $3d\sigma_g$, and $3d\pi_g$ Rydberg states of dinitrogen, 
which belong to the Hopfield series. Our results demonstrate a decrease 
in the decay width for increasing angular momentum and principal quantum 
number within both Rydberg series.
\end{abstract}

%%%%%%%%%%%%%%%%%%%%%%%%%%%%%%%%%%%%%%%%%%%%%%%%%%%%%%%%%%%%%%%%%%%%%
\section{Introduction}
Electronically excited states of atoms and molecules whose energies 
follow the well-known Rydberg formula are called Rydberg states.\cite{reisler_2009} 
They are associated with a decreasing energetic spacing between subsequently 
higher states that converge towards an ionization threshold and are 
characterized by an electron residing in a high-lying and diffuse orbital. 
The Rydberg electron is located far away from the remaining electrons 
and the nuclei; hence, it sees them as an effective ionic core. This 
description motivates the use of the term ``Rydberg state'' for molecules 
and multi-electron atoms, even though it is strictly defined only for 
hydrogen-like one-electron systems. Despite the usefulness of the analogy 
to a hydrogen-like system, this oversimplifies the challenge of describing 
Rydberg states in multi-electron systems, since the screening by the other 
electrons is important. 

Owing to their unique electronic structure, Rydberg states have a plethora 
of fascinating properties. For instance, they are extremely sensitive to 
electric and magnetic fields, making them suitable for applications in 
quantum information science.\cite{Saffman2010,Levine2018} They also serve 
as gateways in reactions involved in plasma environments through dissociative 
recombination\cite{Jensen1999,bardsley_1968} and, furthermore, they play 
a pivotal role in ultrafast spectroscopy and pump-probe 
experiments.\cite{Plunkett2019,moise_2008}

Rydberg states located above the lowest ionization threshold can decay 
through autoionization. In photoionization spectra, the presence of 
autoionizing Rydberg states manifests as sharp features measured near 
higher-lying ionization thresholds.\cite{Carnovale1981,moitra_2021} In electronic-structure 
theory these states are considered electronic Feshbach resonances that 
are embedded in the ionization continuum. Autoionizing Rydberg states 
can be created either by exciting an electron from an inner-valence 
orbital in a neutral closed-shell molecule or by attaching an electron to a 
closed-shell cation.\cite{simons_2022} In this article, we will restrict 
ourselves to Rydberg states that belong to the former case. These states 
decay through a two-electron process, whereby the excited electron is 
emitted and another electron fills the hole in the orbital from which the 
first electron was initially excited. 

Earlier computational studies focusing on autoionization of Rydberg states 
used approaches such as the close-coupling method,\cite{xchem,Puskar_2023} 
R-matrix theory,\cite{rabad_2021} multi-channel quantum defect 
theory,\cite{M_Raoult_1983} and exterior complex scaling combined with 
discrete variable representation.\cite{genevriez_2021,genevriez_2023} 
Overall, however, it is fair to say that the resonance character of Rydberg 
states is often neglected in computational treatments.\cite{reisler_2009} 
On the experimental side, we mention attosecond-noncollinear four-wave-mixing 
spectroscopy,\cite{Puskar_2023} multiphoton Rydberg-dissociation 
spectroscopy,\cite{wehrli_2020} ultrafast XUV/IR pump-probe 
spectroscopy,\cite{Reduzzi_2016} and synchrotron radiation\cite{Schulz_1996,
Codling_1967} that were used to measure decay of autoionizing Rydberg states.

Because resonances are non-stationary states, they cannot be readily 
described by the standard methods used in quantum chemistry that are 
designed for bound states. A pertinent description of resonances would 
therefore involve solving the time-dependent Schr\"odinger equation or 
imposing scattering boundary conditions.\cite{taylor_1972,domcke_1991} 
An alternative approach can, however, be obtained by complex scaling (CS) 
of the Hamiltonian.\cite{aguilar_1971,balslev_1971,reinhardt_1982,
Moiseyev_1998} CS can be viewed as an unbounded similarity transformation 
$\hat{S}H\hat{S}^{-1}$ with an operator $\hat{S}=\text{exp}(i\theta \,r 
\,d/dr)$ that transforms the diverging resonance wave functions into  
$L^2$-integrable eigenfunctions of the complex-scaled Hamiltonian. 
As a result, the Hamiltonian becomes non-Hermitian with complex eigenvalues 
$E=E_R-i\Gamma/2$, known as Siegert energies.\cite{siegert_1939}
The real part of a Siegert energy, $E_R$, describes the position of 
the resonances, while $\Gamma$ is the decay width, which is inversely 
proportional to the lifetime of the state through the energy-time 
uncertainty relation. CS-based approaches present a critical advantage 
in quantum chemistry since they enable resonances to be analyzed in the 
same fashion as bound states. An overview of CS and related non-Hermitian 
methods is available in Refs. \citenum{moiseyev_2011,Jagau_2017,Jagau_2022}. 

Importantly, complex scaling the coordinates of the Hamiltonian by 
$\text{exp}(i\theta)$ is equivalent to scaling the coordinates of 
the wave function by $\text{exp}(-i\theta)$. The latter approach leads 
to the method of complex basis functions (CBFs)\cite{McCurdy_1978} 
and represents an advantage over CS in the context of molecular 
electronic-structure theory since it can be applied to molecules, 
whereas CS is limited to atoms.\cite{Jagau_2017,Jagau_2022} In the 
CBF method, one has the additional flexibility that one can choose 
which basis functions to scale and it has been shown that different 
types of resonances require scaling different parts of a Gaussian basis 
set. For instance, tight functions need to be scaled for describing Auger 
decay,\cite{matz_2022,matz_2023_a} whereas the description of low-lying 
temporary anions requires the scaling of diffuse functions.\cite{white_2015,
white_2015b,white_2017} 

In this article, we aim to investigate autoionizing Rydberg states using 
equation-of-motion coupled-cluster theory in the singles and doubles 
approximation (EOM-CCSD) combined with CS and CBFs. There exist several 
variants of EOM-CC from which excitation energies (EE), electron affinities 
(EA), and ionization potentials (IP) can be obtained\cite{emrich_1981,
stanton_1993,nooijen_1993,stanton_1994,nooijen_1995,krylov_2008,sneskov_2012}. 
The target state is always generated from the CC reference state using an 
operator $\hat{R}$, whose form depends on the EOM-CC variant. In this work, 
we use the excitation energies (EE) variant, where $\hat{R}$ is a spin and 
particle-number conserving excitation operator. Methods combining CS and 
CBFs with EOM-CCSD have already been developed\cite{bravaya_2013,zuev_2014,
white_2017} and were applied to study different types of resonances, 
including core-ionized states involved in Auger decay,\cite{matz_2022,
matz_2023_a,matz_2023_b,jayadev_2023} interatomic and intermolecular 
Couloumbic decay,\cite{parravicini_2023} ionization in static electric 
fields,\cite{jagau_2016,jagau_2018} temporary anions,\cite{white_2017} 
and doubly-excited states of atoms.\cite{bravaya_2013} 

However, CS and CBF methods have so far not been used to describe
autoionizing Ryd\-berg states in Gaussian basis sets. We believe that such 
an approach presents advantages because of its general applicability and 
full consistency because the positions $E_R$ and decay widths $\Gamma$ 
are obtained in the same calculation as real and imaginary parts of the 
Siegert energies. Complex-variable EOM-CC methods can treat bound states 
and resonances on an equal footing together with a rigorous inclusion of 
electron correlation, which is especially important for low-lying Rydberg 
states that are only a few eV above the first ionization energy, since the 
emitted electron is slow and can interact strongly with the remaining 
electrons in the cation.

%%%%%%%%%%%%%%%%%%%%%%%%%%%%%%%%%%%%%%%%%%%%%%%%%%%%%%%%%%%%%%%%%%%%%%

\section{Computational details}
We employ complex-variable EOM-EE-CCSD to study autoionizing Rydberg 
states of neon and dinitrogen in this work. Both of these systems are 
well-studied computationally and experimentally\cite{Josties_2014,
Klar_1994,Min_2008,Codling_1967,Schulz_1996,Puskar_2023,chung_1985,
Huber_1993,woodruff_1977,Warrick_2016,Klinker_2018,M_Raoult_1983,
Little_2013,Reduzzi_2016} and thus represent good model systems to 
investigate the performance of our approach. To the neon atom, both 
CS and CBFs were applied, to dinitrogen only CBFs. A value of 1.099 \AA\
was used for the bond length of dinitrogen. 

%In terms of previous understanding, the same can be said for N$_2$ \cite{chung_1985, Huber_1993,woodruff_1977, Warrick_2016, Klinker_2018, M_Raoult_1983, Little_2013, Reduzzi_2016} as for Neon, yet it represents an additional complexity due to it being a diatomic system. Again, since it is a molecule all calculations carried out on it were done with the method of complex basis functions.\\

The Rydberg states were obtained by solving the EOM-EE-CCSD equations 
in the correlation-consistent Dunning basis sets\cite{woon_1995,
kendall_1992,dunning_1989} aug-cc-pV$X$Z ($X$=T, Q, 5, and 6) further 
augmented by Kaufmann functions.\cite{Kaufmann_1989} Core orbitals were 
frozen in all calculations. To ascertain the positions and widths, scans 
of the complex-scaling angle were carried out between 0$^\circ$ to 45$^\circ$ 
in steps of 1$^\circ$ for CBF calculations and in steps of 0.57$^\circ$ for 
CS calculations. The criterion $\text{min} |dE/d\theta|$ with $E$ referring 
to the complex-valued excitation energy was used to establish the optimal 
angle at which we evaluated the position and width of a given resonance. 
The optimal scaling angles are reported in the Supporting Information.
All calculations were carried out using the Q-Chem software package, 
version 6.0.\cite{qchem}

The Kaufmann basis functions were originally designed to compute energies 
of atomic Rydberg states.\cite{Kaufmann_1989} Their exponents can easily 
be generated using the formula
\begin{equation} \label{eq:kaufmann}
\alpha(n,l)=\Bigl ( \frac{Z}{2n} \Bigr )^2 \frac{1}{(a_{l}n +b_{l})^2} ~, 
\ (n=1,\ 3/2, \ 2, \ 5/2, \ 3, \ ...)
\end{equation}
Here, the exponent $\alpha$ for a certain principal quantum number $n$ 
and angular momentum quantum number $l$ is obtained from pre-calculated 
coefficients $a_l$ and $b_l$ that are tabulated in Ref. \citenum{Kaufmann_1989}. 
$Z$ represents the charge of the ionic core, which is 1 in the case of a 
singly-excited Rydberg state of a neutral species.

Depending on the system and the desired Rydberg state, the required number 
and type of Kaufmann functions are different. Therefore, a large part of 
this article is devoted to the description and analysis of basis-set effects.
For calculations on neon using CS, Kaufmann functions were typically applied 
in the range $n=1.5-3.5$ to describe the Rydberg states with $n=3$. For CBF 
calculations on neon, we employed unscaled Kaufmann functions in the range 
$n=1.0-2.0$ and complex-scaled Kaufmann functions in the range $n=2.0-3.5$. 
In the case of the 4$p$ Rydberg state, Kaufmann functions with up to $n=4.5$ 
were employed. A similar scheme was adopted for dinitrogen with an equal 
number of Kaufmann functions placed at each atom. We employed complex-scaled 
Kaufmann functions with $n=2.5-3.5$ and unscaled Kaufmann functions with $n=2$. 

%%%%%%%%%%%%%%%%%%%%%%%%%%%%%%%%%%%%%%%%%%%%%%%%%%%%%%%%%%%%%%%%%%%%%%%

\section{Results and discussion}

\begin{figure}
\includegraphics[width=0.48\textwidth]{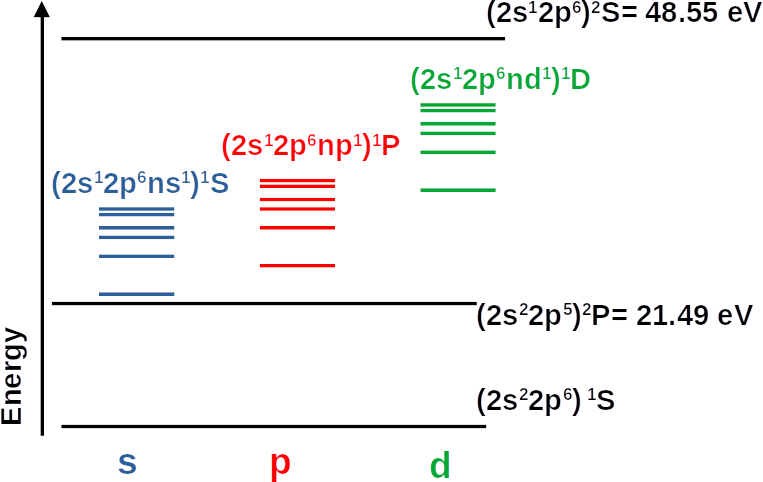} \hspace{0.3cm}
\includegraphics[width=0.48\textwidth]{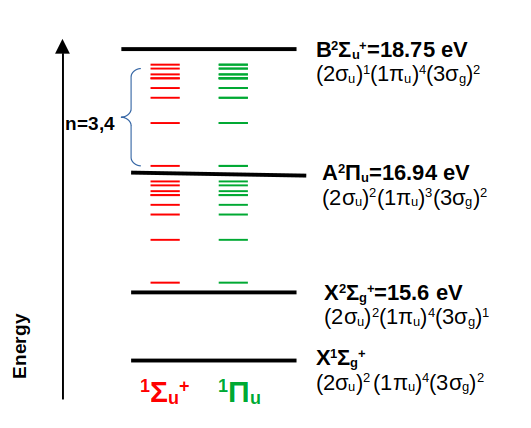}
\caption{Energy level diagrams of Rydberg series in neon and dinitrogen. 
Left: Rydberg series between the first ($^2$P) and second ($^2$S) 
ionization threshold of neon. Ionization energies are obtained by 
EOM-IP-CCSD/aug-cc-pV6Z+6(spd). Right: Rydberg states of $^1\Sigma_u^+$ 
and $^1\Pi_u$ symmetry in the Hopfield series of dinitrogen between 
the second ($A^2\Pi_u$) and third ($B^2\Sigma_u^+$) ionization thresholds. 
Valence electronic configurations are shown for each ionization threshold 
with the full configuration of the ground state being $(1\sigma_g)^2 
(1\sigma_u)^2 (2\sigma_g)^2 (2\sigma_u)^2 (1\pi_u)^4 (3\sigma_g)^2$.
Ionization energies are taken from Ref. \citenum{Reduzzi_2016}.}
\label{fig:ne_n2}
\end{figure}
%WS: Fig. 1a, shouldn't be there ns^1 np^1 nd^1 instead of 3s^1 3p^1 3d^1 labels?

\subsection{Rydberg states of neon}
Exciting an electron from the $2s$ orbital in neon above the first 
ionization threshold at 21.56 eV\cite{kaufman_1972} results in a 
series of autoionizing Rydberg states denoted as $2s^1 2p^6nl$, 
where $n$ is the principal quantum number and $l$ is the angular 
momentum of the Rydberg electron (see Figure \ref{fig:ne_n2}). 
We will restrict ourselves to singlet states with $l= s, p, d$ 
and $n=3, 4$ in the following. For all these states, the only 
open decay channel is the $^2$P ground state of Ne$^+$ at 21.56 eV.
 
According to a recent experiment,\cite{Puskar_2023} the Rydberg 
states are located at 43.67 eV ($3s$), 45.55 eV ($3p$) and 46.97 eV 
($3d$), which means that the electrons emitted in the autoionization 
process have energies of $~$22--26 eV. As documented in the SI, 
our calculations generally overestimate the resonance positions 
somewhat, by ca. 0.05 eV for the $3p$ and the $3d$ state and by 
0.10--0.15 eV for the $3s$ state. For CS-EOMEE-CCSD, the resonance 
positions are higher than those obtained with real-valued EOMEE-CCSD. 
The differences obtained using the best basis set amount to ca. 
0.010 eV for the $3s$ and the $3p$ state and 0.005 eV for the 3d 
state. For CBF-EOMEE-CCSD, the resonance positions are lower than 
the ones from real-valued EOMEE-CCSD with differences of 0.009 eV 
for the $3s$ state and 0.003 eV for the $3p$ state. 

\begin{table} \setlength{\tabcolsep}{7pt}
\centering
\caption{Widths $\Gamma$ in meV of the $3s$ and $3p$ Rydberg states of neon 
computed using CS-EOM-CCSD and CBF-EOM-EE-CCSD with different combinations 
of Dunning and Kaufmann basis sets. The range of $n$ (see Eq. \ref{eq:kaufmann}) 
is reported for each Kaufmann shell. Theoretical and experimental reference 
values are given as well.}
\begin{tabular}{cccr} \hline
Dunning basis & \multicolumn{2}{c}{Kaufmann basis} & $\Gamma$ \\ \hline 
\multicolumn{4}{c}{$3s$ state, CS-EOM-EE-CCSD} \\
cc-pVQZ & \multicolumn{2}{c}{1.0-3.5 ($spd$)} & 143.8 \\
aug-cc-pVQZ & \multicolumn{2}{c}{2.5-3.5 ($spd$)} & 276.0 \\
aug-cc-pV5Z & \multicolumn{2}{c}{none} & 1170.8 \\
aug-cc-pV5Z & \multicolumn{2}{c}{1.5-2.5 ($spd$)} & 110.2 \\
aug-cc-pV5Z & \multicolumn{2}{c}{1.5-3.5 $s$, 2.0-3.5 $p$, 1.5-3.5 $d$} & 135.2 \\
aug-cc-pV5Z & \multicolumn{2}{c}{1.5-3.5 ($spd$)} & 103.6 \\
aug-cc-pV5Z & \multicolumn{2}{c}{1.5-4.0 ($spd$)} & 103.2 \\ 
aug-cc-pV6Z & \multicolumn{2}{c}{1.5-2.5 ($spd$)} & 110.0 \\ 
aug-cc-pV6Z & \multicolumn{2}{c}{1.5-4.0 ($spd$)} & 109.0 \\ \hline
\multicolumn{4}{c}{$3s$ state, CBF-EOM-EE-CCSD} \\
 & unscaled shells & complex-scaled shells & \\
aug-cc-pVQZ & 1.0-1.5 ($sp$), 1.5-2.0 $d$ & 2.0-3.5 ($sp$), 2.5-3.5 $d$ & 69.0 \\
aug-cc-pV5Z & 1.0-1.5 ($sp$), 1.5-2.0 $d$ & 2.0-2.5 ($sp$), 2.5-3.0 $d$ & 128.4 \\
aug-cc-pV5Z & 1.0-1.5 ($sp$), 1.5-2.0 $d$ & 2.0-3.5 ($sp$), 2.5-3.5 $d$ & 109.0 \\
aug-cc-pV6Z & 1.0-1.5 ($sp$), 1.5-2.0 $d$ & 2.0-3.5 ($sp$), 2.5-3.5 $d$ & 110.4 \\ \hline 
\multicolumn{4}{c}{$3s$ state, reference values} \\ 
Experiment\cite{Puskar_2023} &  & \multicolumn{2}{r}{94.0 $\pm$ 27.0} \\
Experiment\cite{Min_2008} &  & \multicolumn{2}{r}{91 $\pm$ 6.0} \\ \hline\hline
\multicolumn{4}{c}{$3p$ state, CS-EOM-EE-CCSD} \\ 
aug-cc-pV5Z & \multicolumn{2}{c}{1.5-2.5 ($spd$)} & 33.2 \\
aug-cc-pV5Z & \multicolumn{2}{c}{1.5-3.5 $s$, 2.0-3.5 $p$, 1.5-3.5 $d$} & 25.4 \\
aug-cc-pV5Z & \multicolumn{2}{c}{1.5-3.5 ($spd$)} & 25.2 \\
aug-cc-pV5Z & \multicolumn{2}{c}{1.5-4.0 ($spd$)} & 24.6 \\
aug-cc-pV6Z & \multicolumn{2}{c}{1.5-4.0 ($spd$)} & 25.0 \\ \hline 
\multicolumn{4}{c}{$3p$ state, CBF-EOM-EE-CCSD} \\
 & unscaled shells & complex-scaled shells & \\
aug-cc-pVQZ & 1.0-1.5 ($sp$), 1.5-2.0 $d$ & 2.0-3.5 ($sp$), 2.5-3.5 $d$ & 6.6 \\
aug-cc-pVQZ & 1.0-1.5 ($sp$), 1.5-2.0 $d$, 2.5 $f$ & 
2.0-3.5 ($sp$), 2.5-3.5 $d$, 3.0-3.5 $f$ & 7.4 \\ 
aug-cc-pV5Z & 1.0-1.5 ($sp$), 1.5-2.0 $d$ & 2.0-2.5 ($sp$), 2.5-3.0 $d$ & 26.6 \\ 
aug-cc-pV5Z & 1.0-1.5 ($sp$), 1.5-2.0 $d$ & 2.0-3.5 ($sp$), 2.5-3.5 $d$ & 29.2 \\
aug-cc-pV6Z & 1.0-1.5 ($sp$), 1.5-2.0 $d$ & 2.0-3.5 ($sp$), 2.5-3.5 $d$ & 29.0 \\ \hline
\multicolumn{4}{c}{$3p$ state, reference values} \\ 
\multicolumn{2}{l}{Close-coupling\cite{Puskar_2023}} &  & 12.4 \\
\multicolumn{2}{l}{R-Matrix+MQDT\cite{Schulz_1996}} &  & 34 \\ 
Experiment\cite{Puskar_2023} &  & \multicolumn{2}{r}{13.7 $\pm$ 2.285} \\ 
Experiment\cite{Schulz_1996} &  & \multicolumn{2}{r}{16.0 $\pm$ 2.0} \\
Experiment\cite{Codling_1967}&  & \multicolumn{2}{r}{13.0 $\pm$ 2.0} \\ \hline
\end{tabular} \label{tab:ne1}
\end{table}

Starting with the CS-EOM-EE-CCSD calculations for the $3s$ state, 
we find that an aug-cc-pVQZ basis set together with 3 $s$, $p$, and 
$d$ shells of Kaufmann functions is the smallest basis set for which 
a minimum in $|dE/d\theta|$ can be found when applying complex scaling 
(see Table \ref{tab:ne1}). This basis yields a width of 276 meV that 
is almost three times as large as the experimental values of 91 and 94 
meV for the width.\cite{Puskar_2023,Min_2008} Using the basis set 
aug-cc-pV5Z without additional Kaufmann functions yields a minimum in 
$|dE/d\theta|$; yet, as shown in Table \ref{tab:ne1}) the agreement 
with the experimental reference values is significantly worse than 
with aug-cc-pVQZ+3(spd). This illustrates the necessity of adding 
Kaufmann functions to obtain accurate values for $\Gamma$. When 
adding the three $s$, $p$, and $d$ shells of Kaufmann functions 
from the aug-cc-pVQZ calculation to the aug-cc-pV5Z basis set, we 
obtain a width of 110 meV, which is in good agreement with recent 
measurements (91 and 94 meV).\cite{Puskar_2023,Min_2008} Adding 
further Kaufmann functions or going to the aug-cc-pV6Z basis does 
not alter the width of the $3s$ Rydberg state significantly and we 
can say that the aug-cc-pV5Z+3(spd) value is converged with respect 
to basis-set size.

Proceeding to the $3p$ Rydberg state, the smallest basis that yields 
a minimum in $|dE/d\theta|$ is aug-cc-pV5Z+3(spd). Also, the agreement 
of our $\Gamma$ values with the experimental values\cite{Puskar_2023,
Codling_1967,Schulz_1996} is somewhat worse compared to the $3s$ state. 
Adding more Kaufmann functions or enlarging the basis set to aug-cc-pV6Z 
does not alter our results significantly. Our best $\Gamma$ value 
of 25 meV is thus converged with respect to the basis set, but still 
differs by ca. 10 meV from the experimentally determined values of 
13--16 meV. In absolute numbers, this deviation is actually smaller 
than the one observed for the $3s$ state, but because the width itself 
is much lower the relative accuracy is better for the $3s$ state than 
for the $3p$ state. We note that the close-coupling method yielded a 
$\Gamma$ value of 12.4 meV,\cite{Puskar_2023} while R-matrix theory 
combined with multi-channel quantum defect theory gave 34 meV.\cite{Schulz_1996}

\begin{table}[tb] \setlength{\tabcolsep}{8pt}
\centering
\caption{Widths $\Gamma$ in meV of the $4p$ and $3d$ Rydberg states of 
neon computed using CS-EOM-CCSD with different combinations of Dunning 
and Kaufmann basis sets. The range of $n$ (see Eq. \ref{eq:kaufmann}) 
is reported for each Kaufmann shell. Theoretical and experimental 
reference values are given as well.}
\begin{tabular}{ccr} \hline
Dunning basis & Kaufmann basis & $\Gamma$ \\ \hline 
\multicolumn{3}{c}{$4p$ state} \\ 
aug-cc-pV5Z & 1.5-4.5 $s$, 2.0-4.5 $p$, 1.5-4.0 $d$ & 13.0 \\
aug-cc-pV5Z & 1.5-4.5 $s$, 2.0-5.0 $p$, 1.5-4.0 $d$ & 12.8 \\
aug-cc-pV6Z & 1.5-4.5 $s$, 2.0-4.0 $p$, 2.0-4.0 $d$ & 11.4 \\ 
Experiment\cite{Codling_1967} & \multicolumn{2}{r}{4.5 $\pm$ 1.5} \\ \hline
\multicolumn{3}{c}{$3d$ state} \\
aug-cc-pV6Z & 1.5-4.0 $s$, 2.0-3.5 $p$, 1.5-4.0 $d$ & 10.0 \\
aug-cc-pV6Z & 1.5-4.0 ($spd$) & 10.8 \\ 
aug-cc-pV6Z & 1.5-4.0 ($spd$), 3.5-4.5 $f$ & 10.4 \\ 
Experiment\cite{Puskar_2023} & \multicolumn{2}{r}{1.54 $\pm$ 0.14} \\ \hline
\end{tabular} \label{tab:ne2}
\end{table}

For the $4p$ state (see Table \ref{tab:ne2}) we observe that 7$s$, 
6$p$, and $6d$ shells of Kaufmann functions on top of aug-cc-pV5Z 
are needed. Adding further basis functions makes little impact so 
that our best value for $\Gamma$ (10.4 meV) overestimates the 
experimental value by 8 meV, which equates to a factor of 2.5.  

The $3d$ Rydberg state poses even more of a challenge since an 
aug-cc-pV6Z basis set together with 6$s$, 5$p$ and 6$d$ shells of 
Kaufmann functions is required to obtain a minimum in $|dE/d\theta|$. 
Our value for the width is between 10.0 and 10.8 meV depending on 
the number of Kaufmann functions added, which is large compared to 
the width of 1.54 meV reported in a recent experimental study.\cite{Puskar_2023}. 
We note that the same authors reported theoretical $\Gamma$ values 
of 0.097 and 1.15 meV. 

Moving on from CS to the CBF method, we find that the smallest possible 
basis for the $3s$ Rydberg state is again aug-cc-pVQZ with an additional 
two and three shells of unscaled and complex-scaled $s$, $p$, and $d$  
Kaufmann functions, respectively. This means that the resulting basis 
set is larger than the smallest one used in the CS calculations. The 
computed widths are in good agreement with those from CS and converge 
to a value of 109 meV that is obtained with the aug-cc-pV5Z+2(spd)+(4s4p3d) 
basis set, where ``2(spd)'' denotes the unscaled Kaufmann functions and 
``(4s4p3d)'' the complex-scaled ones. 

For the $3p$ Rydberg state, the basis set requirements of CBF-EOM-EE-CCSD 
are similar to the $3s$ state, which is in contrast to the CS calculation. 
Surprisingly, for the $3d$ state, none of the tested basis sets produced 
a minimum in $|dE/d\theta|$. As documented in the Supporting information, 
these calculations yield a minimum in $\text{Im}(E)$ at some value of 
$\theta$, but the corresponding widths are clearly unreliable.   

Comparing the different Rydberg states, we observe that the width decreases 
with increasing principal and angular momentum quantum number. The latter 
was also discussed by Puskar \textit{et al.},\cite{Puskar_2023} who related 
it to a lower overlap of orbitals of higher angular momentum with the $2s$ 
hole. Notably, a similar argument about the overlap of the orbitals involved 
in the decay process has also been made to explain trends in Auger decay.\cite{matz_2022} 

%%%%%%%%%%%%%%%%%%%%%%%%%%%%%%%%%%%%%%%%%%%%%%%%%%%%%%%%%%%%%%%%%%%%%%%

\subsection{Rydberg states of dinitrogen}

\begin{table}[tb] \setlength{\tabcolsep}{8pt}
\centering
\caption{Widths $\Gamma$ in meV of Rydberg states of dinitrogen 
computed using CBF-EOM-EE-CCSD with different combinations of Dunning 
and Kaufmann basis sets. The range of $n$ (see Eq. \ref{eq:kaufmann}) 
is reported for each Kaufmann shell. Theoretical and experimental 
reference values are given as well.}
\begin{tabular}{cccc} \hline
 & unscaled & complex-scaled & \\
Dunning basis & Kaufmann basis & Kaufmann basis & $\Gamma$ \\ \hline 
\multicolumn{4}{c}{$3s\sigma_g$ ($^1\Sigma_u^+$)}\\
aug-cc-pVTZ & 2.0 ($spd$) & 2.5-3.5 $s$, 2.5-3.5 $p$, 2.5 $d$ & 117.6 \\
aug-cc-pVQZ & 2.0 ($spd$) & 2.5-3.5 $s$, 2.5-3.5 $p$, 2.5 $d$ & 114.0 \\
\multicolumn{2}{l}{RASSCF+close-coupling\cite{Klinker_2018}} &  & 98.3 \\ \hline
\multicolumn{4}{c}{$4s\sigma_g$ ($^1\Sigma_u^+$)} \\
aug-cc-pVTZ & 2.0 ($spd$) & 2.5-3.5 $s$, 2.5-3.5 $p$, 2.5 $d$ & 77.6 \\
aug-cc-pVQZ & 2.0 ($spd$) & 2.5-3.5 $s$, 2.5-3.5 $p$, 2.5 $d$ & 70.6 \\
\multicolumn{2}{l}{RASSCF+close-coupling\cite{Klinker_2018}} &  & 59.9 \\ \hline
\multicolumn{4}{c}{$3d\sigma_g$ ($^1\Sigma_u^+$)} \\ 
aug-cc-pVTZ & 2.0 ($spd$) & 2.5-3.5 $s$, 2.5-3.5 $p$, 2.5 $d$ & 86.8 \\
aug-cc-pVQZ & 2.0 ($spd$) & 2.5-3.5 $s$, 2.5-3.5 $p$, 2.5 $d$ & 74.4 \\
\multicolumn{2}{l}{RASSCF+close-coupling \cite{Klinker_2018}} &  & 63.3 \\ 
Experiment\cite{Reduzzi_2016} &  &  & 64.5 \\ \hline
\multicolumn{4}{c}{$4d\sigma_g$ ($^1\Sigma_u^+$)} \\ 
aug-cc-pVTZ & 2.0 ($spd$) & 2.5-3.5 $s$, 2.5-3.5 $p$, 2.5 $d$ & 67.6 \\ 
aug-cc-pVQZ & 2.0 ($spd$) & 2.5-3.5 $s$, 2.5-3.5 $p$, 2.5 $d$ & 69.8 \\
\multicolumn{2}{l}{RASSCF+close-coupling\cite{Klinker_2018}} &  & 27.3 \\ \hline
\multicolumn{4}{c}{$3d\pi_g$ ($^1\Pi_u$)} \\
aug-cc-pVTZ & 2.0 ($spd$) & 2.5-3.5 $s$, 2.5-3.5 $p$, 2.5 $d$, 2.5 $f$ & 18.6 \\
\multicolumn{2}{l}{RASSCF+close-coupling\cite{Klinker_2018}} &  & $\approx$ 20 \\
Experiment \cite{Reduzzi_2016} &  &  & 47.7 \\ \hline
\end{tabular} \label{tab:n2_cbf}
\end{table}

For molecular nitrogen, we investigate autoionizing singlet Rydberg 
states between the second (A~$^2 \Pi_u$) and third (B~$^2\Sigma^+_u$) 
ionization thresholds at 16.94 and 18.75 eV.\cite{Reduzzi_2016} These 
states belong to the Hopfield series\cite{hopfield_1930,ogawa_1962} 
and transform according to the $\Sigma^+_u$ and $\Pi_u$ irreducible 
representations. As illustrated in Figure \ref{fig:ne_n2}, states 
of $^1\Sigma^+_u$ symmetry are dominated by transitions from the 
$2\sigma_u$ orbital to Rydberg orbitals of either $d$ or $s$ character 
denoted as $nd\sigma_g$ and $ns\sigma_g$, while states of $^1\Pi_u$ 
symmetry are characterized by transitions from $2\sigma_u$ to Rydberg 
orbitals of $nd\pi_g$ character. Here, $n$ is akin to the principal 
quantum number defined for atoms and previous studies showed that 
the series starts at $n=3$.\cite{M_Raoult_1983,Klinker_2018} 

For all of these states, two decay channels are open, the X~$^2\Sigma_g^+$ 
ground state of N$_2^+$ at 15.6 eV and the A~$^2\Pi_u$ first excited 
state of N$_2^+$ at 16.94 eV. The electrons emitted in the autoionization 
process have energies in the range 0.4--2.6 eV and are thus much slower 
than in the case of neon. 

As shown in Table \ref{tab:n2_cbf}, an aug-cc-pVTZ basis set augmented 
by a single unscaled $s$, $p$ and $d$ shell and $3s$, $3p$, and $1d$ 
complex-scaled shells is needed for all Rydberg states we considered 
to obtain a minimum in $|dE/d\theta|$. We also used the same combination 
of Kaufmann functions with an aug-cc-pVQZ basis set, which gave similar 
results. Only for the $3d\pi_g$ state, a complex-scaled $f$ shell of 
Kaufmann functions is needed. The basis-set requirements of CBF-EOM-EE-CCSD 
are thus considerably less arduous for dinitrogen than for the neon atom.

For the 3$d\sigma_g$ state, our best value for the resonance width 
(74 meV) is in qualitative agreement with the experimental value of 
64.5 meV\cite{Reduzzi_2016} and a previous theoretical one of 63.3 
meV\cite{Klinker_2018} obtained using the XCHEM approach,\cite{xchem} 
which combines restricted active space self-consistent field (RASSCF) 
theory with the close coupling method. In the case of the 3$s\sigma_g$ 
and 4$s\sigma_g$ states, we also find that our widths (114.0 meV and 
70.6 meV, respectively) are in qualitative agreement to the ones 
obtained with XCHEM.\cite{Klinker_2018} In contrast, our width for 
the 4$d\sigma_g$ state (69.8 meV) is significantly larger than the 
XCHEM value (27.3 meV). For the $3d\pi_g$ state, we observe good 
agreement between our result and the XCHEM value (18.6 meV and 
$\approx $20 meV, respectively), but a significant discrepancy with 
experiment (47.7 meV).\cite{Reduzzi_2016} 

Overall, CBF-EOM-EE-CCSD and the XCHEM approach agree that the order 
of the widths is $\Gamma(3s\sigma_g) > \Gamma(3d\sigma_g) > 
\Gamma(4s\sigma_g) > \Gamma(3d\pi_g)$ and disagree only about the 
4$d\sigma_g$ state. We thus observe the same trend as for neon that 
the width decreases with higher angular momentum and higher principal 
quantum number. For instance, the 3$s\sigma_g$ state is ca. 40 meV 
broader than 3$d\sigma_g$ state and the 4$s\sigma_g$ state.

As concerns the real part of the energy, we observe that the differences 
between CBF-EOM-EE-CCSD and real-valued EOM-EE-CCSD do not exceed 
0.04 eV. CBF-EOM-EE-CCSD yields somewhat lower resonance positions 
than real-valued EOM-EE-CCSD for the 3$s\sigma_g$, 4$s\sigma_g$ and 
4$d\sigma_g$ states similar to what we found for neon, whereas the 
opposite trend is seen for the 3$d\sigma_g$ and 3$d\pi_g$ states. 
Our excitation energies are between 17.29 eV and 17.35 eV for the 
3$d\sigma_g$ state and between 17.52 and 17.58 eV for the 3$s\sigma_g$ 
state, which is somewhat larger than the XCHEM values of 17.17 eV 
and 17.39 eV, respectively.\cite{Klinker_2018} However, our energy 
differences between the 3$d\sigma_g$ and 3$s\sigma_g$ states (0.23 eV) 
and between the 4$d\sigma_g$ and 4$s\sigma_g$ states (0.11 eV) are 
in excellent agreement with the XCHEM values. 

% Complex-scaling decreases the positional values of the resonances with 0.01 eV for the 3$s\sigma_g$, 4$s\sigma_g$ and  4$d\sigma_g$ states while it increases with 0.04 eV for the 3$d\sigma_g$ state and 0.03 eV for the 3$d\pi_g$ state.
%The same can be observed for the relative energy position between the 4$d\sigma_g$ and 4$s\sigma_g$ states, which is 0.110-0.118 eV compared to 0.106 eV.

%%%%%%%%%%%%%%%%%%%%%%%%%%%%%%%%%%%%%%%%%%%%%%%%%%%%%%%%%%%%%%%%%%%%%%%%%

\section{Conclusions}

To conclude, we have demonstrated that CS-EOM-EE-CCSD and CBF-EOM-EE-CCSD 
can be used to compute decay widths of atomic and molecular Rydberg states. 
For both investigated systems, neon and dinitrogen, we observe that the 
width decreases with higher angular momentum and principal quantum number 
of the Rydberg electron, which confirms earlier studies. 

The autoionization process differs considerably between neon and dinitrogen: 
For the $2s^12p^6nl$ states of neon, the emitted electron has an energy of 
more than 20 eV and is thus considerably faster than in the case of dinitrogen 
($<$3 eV). Also, there is only one decay channel open in neon as compared to 
two decay channels in dinitrogen. 

We showed that Kaufmann basis functions, which were originally designed 
to compute energies of atomic Rydberg states are adequate to compute decay 
widths as well and that they can be employed for computations on atoms 
and molecules alike. As the resulting basis sets are quite large, it is 
clear that these computations are expensive, but we note that the 
complex-variable treatment does not always require a larger basis set 
than the corresponding real-valued computation. 

For neon, this appears to be the case, since the smallest adequate basis 
is aug-cc-pVQZ+3(spd) in the complex-variable treatment, while a smaller 
basis is sufficient for accurate energies. In contrast, already the 
aug-cc-pVTZ+4s4p2d basis is adequate for computing decay width of Rydberg 
states of dinitrogen. The different minimum requirements for neon and 
dinitrogen may be due to the different speed of the emitted electron or 
the different number of open channels but it could also reflect the simple 
fact that the basis set is effectively twice as large if there are two 
atoms instead of one. 

In any case, consider the relatively modest basis-set requirements of 
the Rydberg states of dinitrogen as an encouragement to apply our method 
to autoionizing states in larger organic molecules, where there is 
substantial mixing of Rydberg and valence states. We also consider 
it worthwhile to extend methods for the computation of partial decay 
widths within complex-variable treatments\cite{matz_2022,matz_2023_a} 
to Rydberg states to investigate the relevance of the different decay 
channels. Finally, a further prospect that would be interesting to 
pursue is the visualization of autoionizing Rydberg states and their 
decay in terms of natural transition orbitals\cite{krylov_2020} and 
the recently introduced natural Auger orbitals.\cite{jayadev_2023_b} 

% Furthermore, this cost increases with angular momentum and principal quantum number in CS. This is, however, not the case in CBF, where the basis set is similar for 3s and 3p calculations. Still. the minimal basis set is large. 
% since in future studies we expect to be able to study larger systems such as benzene and ethylene. 
% This would be helpful to assign full symmetry of Rydberg states obtained in standard quantum-molecular calculations.

\begin{acknowledgement}
This work has been supported by the Central Europe Leuven Strategic 
Alliance (CELSA) (grant 22/014 to W.S. and T.C.J.) and by the European 
Research Council (ERC) under the European Union's Horizon 2020 research 
and innovation program (grant 851766 to T.C.J.)
\end{acknowledgement}

\begin{suppinfo}
The Supporting Information is available free of charge at \texttt{<link 
to be inserted>}: Energies, decay widths, and optimal complex-scaling 
angles of all studied Rydberg states computed with different basis sets; 
exemplary Q-Chem inputs for CS-EOMEE-CCSD and CBF-EOMEE-CCSD calculations.
\end{suppinfo}

\bibliography{achemso-demo}

\providecommand{\latin}[1]{#1}
\makeatletter
\providecommand{\doi}
  {\begingroup\let\do\@makeother\dospecials
  \catcode`\{=1 \catcode`\}=2 \doi@aux}
\providecommand{\doi@aux}[1]{\endgroup\texttt{#1}}
\makeatother
\providecommand*\mcitethebibliography{\thebibliography}
\csname @ifundefined\endcsname{endmcitethebibliography}
  {\let\endmcitethebibliography\endthebibliography}{}
\begin{mcitethebibliography}{70}
\providecommand*\natexlab[1]{#1}
\providecommand*\mciteSetBstSublistMode[1]{}
\providecommand*\mciteSetBstMaxWidthForm[2]{}
\providecommand*\mciteBstWouldAddEndPuncttrue
  {\def\EndOfBibitem{\unskip.}}
\providecommand*\mciteBstWouldAddEndPunctfalse
  {\let\EndOfBibitem\relax}
\providecommand*\mciteSetBstMidEndSepPunct[3]{}
\providecommand*\mciteSetBstSublistLabelBeginEnd[3]{}
\providecommand*\EndOfBibitem{}
\mciteSetBstSublistMode{f}
\mciteSetBstMaxWidthForm{subitem}{(\alph{mcitesubitemcount})}
\mciteSetBstSublistLabelBeginEnd
  {\mcitemaxwidthsubitemform\space}
  {\relax}
  {\relax}

\bibitem[Reisler and Krylov(2009)Reisler, and Krylov]{reisler_2009}
Reisler,~H.; Krylov,~A.~I. Interacting Rydberg and valence states in radicals
  and molecules: experimental and theoretical studies. \emph{Int. Rev. Phys.
  Chem.} \textbf{2009}, \emph{28}, 267--308\relax
\mciteBstWouldAddEndPuncttrue
\mciteSetBstMidEndSepPunct{\mcitedefaultmidpunct}
{\mcitedefaultendpunct}{\mcitedefaultseppunct}\relax
\EndOfBibitem
\bibitem[Saffman \latin{et~al.}(2010)Saffman, Walker, and
  M\o{}lmer]{Saffman2010}
Saffman,~M.; Walker,~T.~G.; M\o{}lmer,~K. Quantum information with Rydberg
  atoms. \emph{Rev. Mod. Phys.} \textbf{2010}, \emph{82}, 2313--2363\relax
\mciteBstWouldAddEndPuncttrue
\mciteSetBstMidEndSepPunct{\mcitedefaultmidpunct}
{\mcitedefaultendpunct}{\mcitedefaultseppunct}\relax
\EndOfBibitem
\bibitem[Levine \latin{et~al.}(2018)Levine, Keesling, Omran, Bernien, Schwartz,
  Zibrov, Endres, Greiner, Vuleti\'c, and Lukin]{Levine2018}
Levine,~H.; Keesling,~A.; Omran,~A.; Bernien,~H.; Schwartz,~S.; Zibrov,~A.~S.;
  Endres,~M.; Greiner,~M.; Vuleti\'c,~V.; Lukin,~M.~D. High-Fidelity Control
  and Entanglement of Rydberg-Atom Qubits. \emph{Phys. Rev. Lett.}
  \textbf{2018}, \emph{121}, 123603\relax
\mciteBstWouldAddEndPuncttrue
\mciteSetBstMidEndSepPunct{\mcitedefaultmidpunct}
{\mcitedefaultendpunct}{\mcitedefaultseppunct}\relax
\EndOfBibitem
\bibitem[Jensen \latin{et~al.}(1999)Jensen, Bilodeau, Heber, Pedersen, Safvan,
  Urbain, Zajfman, and Andersen]{Jensen1999}
Jensen,~M.~J.; Bilodeau,~R.~C.; Heber,~O.; Pedersen,~H.~B.; Safvan,~C.~P.;
  Urbain,~X.; Zajfman,~D.; Andersen,~L.~H. Dissociative recombination and
  excitation of {H$_2$O$^+$} and {HDO$^+$}. \emph{Phys. Rev. A} \textbf{1999},
  \emph{60}, 2970--2976\relax
\mciteBstWouldAddEndPuncttrue
\mciteSetBstMidEndSepPunct{\mcitedefaultmidpunct}
{\mcitedefaultendpunct}{\mcitedefaultseppunct}\relax
\EndOfBibitem
\bibitem[Bardsley(1968)]{bardsley_1968}
Bardsley,~J.~N. The theory of dissociative recombination. \emph{J. Phys. B: At.
  Mol. Opt. Phys.} \textbf{1968}, \emph{1}, 365\relax
\mciteBstWouldAddEndPuncttrue
\mciteSetBstMidEndSepPunct{\mcitedefaultmidpunct}
{\mcitedefaultendpunct}{\mcitedefaultseppunct}\relax
\EndOfBibitem
\bibitem[Plunkett \latin{et~al.}(2019)Plunkett, Harkema, Lucchese, McCurdy, and
  Sandhu]{Plunkett2019}
Plunkett,~A.; Harkema,~N.; Lucchese,~R.~R.; McCurdy,~C.~W.; Sandhu,~A.
  Ultrafast Rydberg-state dissociation in oxygen: Identifying the role of
  multielectron excitations. \emph{Phys. Rev. A} \textbf{2019}, \emph{99},
  063403\relax
\mciteBstWouldAddEndPuncttrue
\mciteSetBstMidEndSepPunct{\mcitedefaultmidpunct}
{\mcitedefaultendpunct}{\mcitedefaultseppunct}\relax
\EndOfBibitem
\bibitem[Moise \latin{et~al.}(2008)Moise, Alagia, Banchi, Ferianis, Prince, and
  Richter]{moise_2008}
Moise,~A.; Alagia,~M.; Banchi,~L.; Ferianis,~M.; Prince,~K.~C.; Richter,~R.
  Pump-probe studies of autoionizing states of noble gases combining laser and
  synchrotron radiation--The nf' Rydberg states of neon. \emph{Nucl. Instrum.
  Methods Phys. Res. A} \textbf{2008}, \emph{588}, 502--508\relax
\mciteBstWouldAddEndPuncttrue
\mciteSetBstMidEndSepPunct{\mcitedefaultmidpunct}
{\mcitedefaultendpunct}{\mcitedefaultseppunct}\relax
\EndOfBibitem
\bibitem[Carnovale \latin{et~al.}(1981)Carnovale, White, and
  Brion]{Carnovale1981}
Carnovale,~F.; White,~M.; Brion,~C. Absolute dipole oscillator strengths for
  photoabsorption and photoionization of carbon disulphide. \emph{J. Electron
  Spectrosc. and Relat. Phenom.} \textbf{1981}, \emph{24}, 63--76\relax
\mciteBstWouldAddEndPuncttrue
\mciteSetBstMidEndSepPunct{\mcitedefaultmidpunct}
{\mcitedefaultendpunct}{\mcitedefaultseppunct}\relax
\EndOfBibitem
\bibitem[Moitra \latin{et~al.}(2021)Moitra, Coriani, and Decleva]{moitra_2021}
Moitra,~T.; Coriani,~S.; Decleva,~P. Capturing correlation effects on
  photoionization dynamics. \emph{J. Chem Theory Comput.} \textbf{2021},
  \emph{17}, 5064--5079\relax
\mciteBstWouldAddEndPuncttrue
\mciteSetBstMidEndSepPunct{\mcitedefaultmidpunct}
{\mcitedefaultendpunct}{\mcitedefaultseppunct}\relax
\EndOfBibitem
\bibitem[Simons(2022)]{simons_2022}
Simons,~J. {Do not forget the Rydberg orbitals}. \emph{J. Chem. Phys.}
  \textbf{2022}, \emph{156}, 100901\relax
\mciteBstWouldAddEndPuncttrue
\mciteSetBstMidEndSepPunct{\mcitedefaultmidpunct}
{\mcitedefaultendpunct}{\mcitedefaultseppunct}\relax
\EndOfBibitem
\bibitem[Marante \latin{et~al.}(2017)Marante, Klinker, Corral,
  González-Vázquez, Argenti, and Martín]{xchem}
Marante,~C.; Klinker,~M.; Corral,~I.; González-Vázquez,~J.; Argenti,~L.;
  Martín,~F. Hybrid-Basis Close-Coupling Interface to Quantum Chemistry
  Packages for the Treatment of Ionization Problems. \emph{J. Chem. Theory
  Comput.} \textbf{2017}, \emph{13}, 499--514\relax
\mciteBstWouldAddEndPuncttrue
\mciteSetBstMidEndSepPunct{\mcitedefaultmidpunct}
{\mcitedefaultendpunct}{\mcitedefaultseppunct}\relax
\EndOfBibitem
\bibitem[Puskar \latin{et~al.}(2023)Puskar, Lin, Gaynor, Schuchter,
  Chattopadhyay, Marante, Fidler, Keenan, Argenti, Neumark, and
  Leone]{Puskar_2023}
Puskar,~N.~G.; Lin,~Y.-C.; Gaynor,~J.~D.; Schuchter,~M.~C.; Chattopadhyay,~S.;
  Marante,~C.; Fidler,~A.~P.; Keenan,~C.~L.; Argenti,~L.; Neumark,~D.~M.;
  Leone,~S.~R. Measuring autoionization decay lifetimes of optically forbidden
  inner valence excited states in neon atoms with attosecond noncollinear
  four-wave-mixing spectroscopy. \emph{Phys. Rev. A} \textbf{2023}, \emph{107},
  033117\relax
\mciteBstWouldAddEndPuncttrue
\mciteSetBstMidEndSepPunct{\mcitedefaultmidpunct}
{\mcitedefaultendpunct}{\mcitedefaultseppunct}\relax
\EndOfBibitem
\bibitem[Rabad\'an and Gorfinkiel(2021)Rabad\'an, and Gorfinkiel]{rabad_2021}
Rabad\'an,~I.; Gorfinkiel,~J.~D. Toward a description of electron-induced
  dissociative excitation in {H$_2$O$^+$}: Investigation of three resonances
  above the {$\tilde{B}$} state. \emph{Phys. Rev. A} \textbf{2021}, \emph{103},
  032804\relax
\mciteBstWouldAddEndPuncttrue
\mciteSetBstMidEndSepPunct{\mcitedefaultmidpunct}
{\mcitedefaultendpunct}{\mcitedefaultseppunct}\relax
\EndOfBibitem
\bibitem[Raoult \latin{et~al.}(1983)Raoult, Rouzo, Raseev, and
  Lefebvre-Brion]{M_Raoult_1983}
Raoult,~M.; Rouzo,~H.~L.; Raseev,~G.; Lefebvre-Brion,~H. Ab initio approach to
  the multichannel quantum defect calculation of the electronic autoionisation
  in the Hopfield series of N$_2$. \emph{J. Phys. B: At. Mol.} \textbf{1983},
  \emph{16}, 4601\relax
\mciteBstWouldAddEndPuncttrue
\mciteSetBstMidEndSepPunct{\mcitedefaultmidpunct}
{\mcitedefaultendpunct}{\mcitedefaultseppunct}\relax
\EndOfBibitem
\bibitem[G\'en\'evriez(2021)]{genevriez_2021}
G\'en\'evriez,~M. Theoretical approaches for doubly-excited {Rydberg} states in
  quasi-two-electron systems: two-electron dynamics far away from the nucleus.
  \emph{Mol. Phys.} \textbf{2021}, \emph{119}, e1861353\relax
\mciteBstWouldAddEndPuncttrue
\mciteSetBstMidEndSepPunct{\mcitedefaultmidpunct}
{\mcitedefaultendpunct}{\mcitedefaultseppunct}\relax
\EndOfBibitem
\bibitem[G\'en\'evriez and Eichmann(2023)G\'en\'evriez, and
  Eichmann]{genevriez_2023}
G\'en\'evriez,~M.; Eichmann,~U. Isolated-core quadrupole excitation of highly
  excited autoionizing {Rydberg} states. \emph{Phys. Rev. A} \textbf{2023},
  \emph{107}, 012817\relax
\mciteBstWouldAddEndPuncttrue
\mciteSetBstMidEndSepPunct{\mcitedefaultmidpunct}
{\mcitedefaultendpunct}{\mcitedefaultseppunct}\relax
\EndOfBibitem
\bibitem[Wehrli \latin{et~al.}(2020)Wehrli, G\'en\'evriez, Knecht, Reiher, and
  Merkt]{wehrli_2020}
Wehrli,~D.; G\'en\'evriez,~M.; Knecht,~S.; Reiher,~M.; Merkt,~F. Complete
  characterization of the 3p {Rydberg} complex of a molecular ion: {MgAr$^+$}.
  {I}. Observation of the {Mg(3p$_\sigma$)Ar$^+$ B$^+$} state and determination
  of its structure and dynamics. \emph{J. Chem. Phys.} \textbf{2020},
  \emph{153}, 074310\relax
\mciteBstWouldAddEndPuncttrue
\mciteSetBstMidEndSepPunct{\mcitedefaultmidpunct}
{\mcitedefaultendpunct}{\mcitedefaultseppunct}\relax
\EndOfBibitem
\bibitem[Reduzzi \latin{et~al.}(2016)Reduzzi, Chu, Feng, Dubrouil, Hummert,
  Calegari, Frassetto, Poletto, Kornilov, Nisoli, Lin, and
  Sansone]{Reduzzi_2016}
Reduzzi,~M.; Chu,~W.-C.; Feng,~C.; Dubrouil,~A.; Hummert,~J.; Calegari,~F.;
  Frassetto,~F.; Poletto,~L.; Kornilov,~O.; Nisoli,~M.; Lin,~C.-D.; Sansone,~G.
  Observation of autoionization dynamics and sub-cycle quantum beating in
  electronic molecular wave packets. \emph{J. Phys. B: At. Mol. Opt. Phys.}
  \textbf{2016}, \emph{49}, 065102\relax
\mciteBstWouldAddEndPuncttrue
\mciteSetBstMidEndSepPunct{\mcitedefaultmidpunct}
{\mcitedefaultendpunct}{\mcitedefaultseppunct}\relax
\EndOfBibitem
\bibitem[Schulz \latin{et~al.}(1996)Schulz, Domke, P\"uttner, Guti\'errez,
  Kaindl, Miecznik, and Greene]{Schulz_1996}
Schulz,~K.; Domke,~M.; P\"uttner,~R.; Guti\'errez,~A.; Kaindl,~G.;
  Miecznik,~G.; Greene,~C.~H. High-resolution experimental and theoretical
  study of singly and doubly excited resonances in ground-state photoionization
  of neon. \emph{Phys. Rev. A} \textbf{1996}, \emph{54}, 3095--3112\relax
\mciteBstWouldAddEndPuncttrue
\mciteSetBstMidEndSepPunct{\mcitedefaultmidpunct}
{\mcitedefaultendpunct}{\mcitedefaultseppunct}\relax
\EndOfBibitem
\bibitem[Codling \latin{et~al.}(1967)Codling, Madden, and Ederer]{Codling_1967}
Codling,~K.; Madden,~R.~P.; Ederer,~D.~L. Resonances in the Photo-Ionization
  Continuum of Ne I (20-150 eV). \emph{Phys. Rev.} \textbf{1967}, \emph{155},
  26--37\relax
\mciteBstWouldAddEndPuncttrue
\mciteSetBstMidEndSepPunct{\mcitedefaultmidpunct}
{\mcitedefaultendpunct}{\mcitedefaultseppunct}\relax
\EndOfBibitem
\bibitem[Taylor(1972)]{taylor_1972}
Taylor,~J.~R. \emph{{S}cattering {T}heory: {T}he quantum {T}heory on
  {N}onrelativistic {C}ollisions}; Wiley, New York, 1972\relax
\mciteBstWouldAddEndPuncttrue
\mciteSetBstMidEndSepPunct{\mcitedefaultmidpunct}
{\mcitedefaultendpunct}{\mcitedefaultseppunct}\relax
\EndOfBibitem
\bibitem[Domcke(1991)]{domcke_1991}
Domcke,~W. Theory of resonance and threshold effects in electron-molecule
  collisions: The projection-operator approach. \emph{Phys. Rep.}
  \textbf{1991}, \emph{208}, 97--188\relax
\mciteBstWouldAddEndPuncttrue
\mciteSetBstMidEndSepPunct{\mcitedefaultmidpunct}
{\mcitedefaultendpunct}{\mcitedefaultseppunct}\relax
\EndOfBibitem
\bibitem[Aguilar and Combes(1971)Aguilar, and Combes]{aguilar_1971}
Aguilar,~J.; Combes,~J.~M. A Class of Analytic Perturbations for One-body
  Schr\"odinger Hamiltonians. \emph{Commun. Math. Phys.} \textbf{1971},
  \emph{22}, 269--279\relax
\mciteBstWouldAddEndPuncttrue
\mciteSetBstMidEndSepPunct{\mcitedefaultmidpunct}
{\mcitedefaultendpunct}{\mcitedefaultseppunct}\relax
\EndOfBibitem
\bibitem[Balslev and Combes(1971)Balslev, and Combes]{balslev_1971}
Balslev,~E.; Combes,~J.~M. Spectral Properties of Many-body Schr\"odinger
  Operators with Dilatation-analytic Interactions. \emph{Commun. Math. Phys.}
  \textbf{1971}, \emph{22}, 280--294\relax
\mciteBstWouldAddEndPuncttrue
\mciteSetBstMidEndSepPunct{\mcitedefaultmidpunct}
{\mcitedefaultendpunct}{\mcitedefaultseppunct}\relax
\EndOfBibitem
\bibitem[Reinhardt(1982)]{reinhardt_1982}
Reinhardt,~W.~P. Complex Coordinates in the Theory of Atomic and Molecular
  Structure and Dynamics. \emph{Annu. Rev. Phys. Chem.} \textbf{1982},
  \emph{33}, 223--255\relax
\mciteBstWouldAddEndPuncttrue
\mciteSetBstMidEndSepPunct{\mcitedefaultmidpunct}
{\mcitedefaultendpunct}{\mcitedefaultseppunct}\relax
\EndOfBibitem
\bibitem[Moiseyev(1998)]{Moiseyev_1998}
Moiseyev,~N. Quantum theory of resonances: calculating energies, widths and
  cross-sections by complex scaling. \emph{Phys. Rep.} \textbf{1998},
  \emph{302}, 212--293\relax
\mciteBstWouldAddEndPuncttrue
\mciteSetBstMidEndSepPunct{\mcitedefaultmidpunct}
{\mcitedefaultendpunct}{\mcitedefaultseppunct}\relax
\EndOfBibitem
\bibitem[Siegert(1939)]{siegert_1939}
Siegert,~A. J.~F. On the Derivation of the Dispersion Formula for Nuclear
  Reactions. \emph{Phys. Rev.} \textbf{1939}, \emph{56}, 750--752\relax
\mciteBstWouldAddEndPuncttrue
\mciteSetBstMidEndSepPunct{\mcitedefaultmidpunct}
{\mcitedefaultendpunct}{\mcitedefaultseppunct}\relax
\EndOfBibitem
\bibitem[Moiseyev(2011)]{moiseyev_2011}
Moiseyev,~N. \emph{Non-Hermitian Quantum Mechanics}; Cambridge University
  Press, 2011\relax
\mciteBstWouldAddEndPuncttrue
\mciteSetBstMidEndSepPunct{\mcitedefaultmidpunct}
{\mcitedefaultendpunct}{\mcitedefaultseppunct}\relax
\EndOfBibitem
\bibitem[Jagau \latin{et~al.}(2017)Jagau, Bravaya, and Krylov]{Jagau_2017}
Jagau,~T.-C.; Bravaya,~K.~B.; Krylov,~A.~I. Extending Quantum Chemistry of
  Bound States to Electronic Resonances. \emph{Annu. Rev. Phys. Chem.}
  \textbf{2017}, \emph{68}, 525--553\relax
\mciteBstWouldAddEndPuncttrue
\mciteSetBstMidEndSepPunct{\mcitedefaultmidpunct}
{\mcitedefaultendpunct}{\mcitedefaultseppunct}\relax
\EndOfBibitem
\bibitem[Jagau(2022)]{Jagau_2022}
Jagau,~T.-C. Theory of electronic resonances: fundamental aspects and recent
  advances. \emph{Chem. Commun.} \textbf{2022}, \emph{58}, 5205--5224\relax
\mciteBstWouldAddEndPuncttrue
\mciteSetBstMidEndSepPunct{\mcitedefaultmidpunct}
{\mcitedefaultendpunct}{\mcitedefaultseppunct}\relax
\EndOfBibitem
\bibitem[McCurdy and Rescigno(1978)McCurdy, and Rescigno]{McCurdy_1978}
McCurdy,~C.~W.; Rescigno,~T.~N. Extension of the Method of Complex Basis
  Functions to Molecular Resonances. \emph{Phys. Rev. Lett.} \textbf{1978},
  \emph{41}, 1364--1368\relax
\mciteBstWouldAddEndPuncttrue
\mciteSetBstMidEndSepPunct{\mcitedefaultmidpunct}
{\mcitedefaultendpunct}{\mcitedefaultseppunct}\relax
\EndOfBibitem
\bibitem[Matz and Jagau(2022)Matz, and Jagau]{matz_2022}
Matz,~F.; Jagau,~T.-C. Molecular Auger decay rates from complex-variable
  coupled-cluster theory. \emph{J. Chem. Phys.} \textbf{2022}, \emph{156},
  114117\relax
\mciteBstWouldAddEndPuncttrue
\mciteSetBstMidEndSepPunct{\mcitedefaultmidpunct}
{\mcitedefaultendpunct}{\mcitedefaultseppunct}\relax
\EndOfBibitem
\bibitem[Matz and Jagau(2023)Matz, and Jagau]{matz_2023_a}
Matz,~F.; Jagau,~T.-C. Channel-specific core-valence projectors for determining
  partial Auger decay widths. \emph{Mol. Phys.} \textbf{2023}, \emph{121},
  e2105270\relax
\mciteBstWouldAddEndPuncttrue
\mciteSetBstMidEndSepPunct{\mcitedefaultmidpunct}
{\mcitedefaultendpunct}{\mcitedefaultseppunct}\relax
\EndOfBibitem
\bibitem[White \latin{et~al.}(2015)White, Head-Gordon, and
  {McCurdy}]{white_2015}
White,~A.~F.; Head-Gordon,~M.; {McCurdy},~C.~W. Complex basis functions
  revisited: Implementation with applications to carbon tetrafluoride and
  aromatic N-containing heterocycles within the static-exchange approximation.
  \emph{J. Chem. Phys.} \textbf{2015}, \emph{142}, 054103\relax
\mciteBstWouldAddEndPuncttrue
\mciteSetBstMidEndSepPunct{\mcitedefaultmidpunct}
{\mcitedefaultendpunct}{\mcitedefaultseppunct}\relax
\EndOfBibitem
\bibitem[White \latin{et~al.}(2015)White, {McCurdy}, and
  Head-Gordon]{white_2015b}
White,~A.~F.; {McCurdy},~C.~W.; Head-Gordon,~M. Restricted and unrestricted
  non-Hermitian Hartree-Fock: Theory, practical considerations, and
  applications to metastable molecular anions. \emph{J. Chem. Phys.}
  \textbf{2015}, \emph{143}, 074103\relax
\mciteBstWouldAddEndPuncttrue
\mciteSetBstMidEndSepPunct{\mcitedefaultmidpunct}
{\mcitedefaultendpunct}{\mcitedefaultseppunct}\relax
\EndOfBibitem
\bibitem[White \latin{et~al.}(2017)White, Epifanovsky, McCurdy, and
  Head-Gordon]{white_2017}
White,~A.~F.; Epifanovsky,~E.; McCurdy,~C.~W.; Head-Gordon,~M. Second order
  Møller-Plesset and coupled cluster singles and doubles methods with complex
  basis functions for resonances in electron-molecule scattering. \emph{J.
  Chem. Phys.} \textbf{2017}, \emph{146}, 234107\relax
\mciteBstWouldAddEndPuncttrue
\mciteSetBstMidEndSepPunct{\mcitedefaultmidpunct}
{\mcitedefaultendpunct}{\mcitedefaultseppunct}\relax
\EndOfBibitem
\bibitem[Emrich(1981)]{emrich_1981}
Emrich,~K. An extension of the coupled cluster formalism to excited states (I).
  \emph{Nucl. Phys. A} \textbf{1981}, \emph{351}, 379--396\relax
\mciteBstWouldAddEndPuncttrue
\mciteSetBstMidEndSepPunct{\mcitedefaultmidpunct}
{\mcitedefaultendpunct}{\mcitedefaultseppunct}\relax
\EndOfBibitem
\bibitem[Stanton and Bartlett(1993)Stanton, and Bartlett]{stanton_1993}
Stanton,~J.~F.; Bartlett,~R.~J. The equation of motion coupled‐cluster
  method. A systematic biorthogonal approach to molecular excitation energies,
  transition probabilities, and excited state properties. \emph{J. Chem. Phys.}
  \textbf{1993}, \emph{98}, 7029--7039\relax
\mciteBstWouldAddEndPuncttrue
\mciteSetBstMidEndSepPunct{\mcitedefaultmidpunct}
{\mcitedefaultendpunct}{\mcitedefaultseppunct}\relax
\EndOfBibitem
\bibitem[Nooijen and Snijders(1993)Nooijen, and Snijders]{nooijen_1993}
Nooijen,~M.; Snijders,~J.~G. Coupled cluster Green's function method: Working
  equations and applications. \emph{Int. J. Quantum Chem.} \textbf{1993},
  \emph{48}, 15--48\relax
\mciteBstWouldAddEndPuncttrue
\mciteSetBstMidEndSepPunct{\mcitedefaultmidpunct}
{\mcitedefaultendpunct}{\mcitedefaultseppunct}\relax
\EndOfBibitem
\bibitem[Stanton and Gauss(1994)Stanton, and Gauss]{stanton_1994}
Stanton,~J.~F.; Gauss,~J. Analytic energy derivatives for ionized states
  described by the equation‐of‐motion coupled cluster method. \emph{J.
  Chem. Phys.} \textbf{1994}, \emph{101}, 8938--8944\relax
\mciteBstWouldAddEndPuncttrue
\mciteSetBstMidEndSepPunct{\mcitedefaultmidpunct}
{\mcitedefaultendpunct}{\mcitedefaultseppunct}\relax
\EndOfBibitem
\bibitem[Nooijen and Bartlett(1995)Nooijen, and Bartlett]{nooijen_1995}
Nooijen,~M.; Bartlett,~R.~J. Equation of motion coupled cluster method for
  electron attachment. \emph{J. Chem. Phys.} \textbf{1995}, \emph{102},
  3629--3647\relax
\mciteBstWouldAddEndPuncttrue
\mciteSetBstMidEndSepPunct{\mcitedefaultmidpunct}
{\mcitedefaultendpunct}{\mcitedefaultseppunct}\relax
\EndOfBibitem
\bibitem[Krylov(2008)]{krylov_2008}
Krylov,~A.~I. Equation-of-Motion Coupled-Cluster Methods for Open-Shell and
  Electronically Excited Species: The Hitchhiker's Guide to Fock Space.
  \emph{Annu. Rev. Phys. Chem.} \textbf{2008}, \emph{59}, 433--462\relax
\mciteBstWouldAddEndPuncttrue
\mciteSetBstMidEndSepPunct{\mcitedefaultmidpunct}
{\mcitedefaultendpunct}{\mcitedefaultseppunct}\relax
\EndOfBibitem
\bibitem[Sneskov and Christiansen(2012)Sneskov, and Christiansen]{sneskov_2012}
Sneskov,~K.; Christiansen,~O. Excited state coupled cluster methods.
  \emph{WIREs Comput. Mol. Sci.} \textbf{2012}, \emph{2}, 566--584\relax
\mciteBstWouldAddEndPuncttrue
\mciteSetBstMidEndSepPunct{\mcitedefaultmidpunct}
{\mcitedefaultendpunct}{\mcitedefaultseppunct}\relax
\EndOfBibitem
\bibitem[Bravaya \latin{et~al.}(2013)Bravaya, Zuev, Epifanovsky, and
  Krylov]{bravaya_2013}
Bravaya,~K.~B.; Zuev,~D.; Epifanovsky,~E.; Krylov,~A.~I. Complex-scaled
  equation-of-motion coupled-cluster method with single and double
  substitutions for autoionizing excited states: Theory, implementation, and
  examples. \emph{J. Chem. Phys.} \textbf{2013}, \emph{138}, 124106\relax
\mciteBstWouldAddEndPuncttrue
\mciteSetBstMidEndSepPunct{\mcitedefaultmidpunct}
{\mcitedefaultendpunct}{\mcitedefaultseppunct}\relax
\EndOfBibitem
\bibitem[Zuev \latin{et~al.}(2014)Zuev, Jagau, Bravaya, Epifanovsky, Shao,
  Sundstrom, Head-Gordon, and Krylov]{zuev_2014}
Zuev,~D.; Jagau,~T.-C.; Bravaya,~K.~B.; Epifanovsky,~E.; Shao,~Y.;
  Sundstrom,~E.; Head-Gordon,~M.; Krylov,~A.~I. Complex absorbing potentials
  within EOM-CC family of methods: Theory, implementation, and benchmarks.
  \emph{J. Chem. Phys.} \textbf{2014}, \emph{141}, 024102\relax
\mciteBstWouldAddEndPuncttrue
\mciteSetBstMidEndSepPunct{\mcitedefaultmidpunct}
{\mcitedefaultendpunct}{\mcitedefaultseppunct}\relax
\EndOfBibitem
\bibitem[Matz \latin{et~al.}(2023)Matz, Nijssen, and Jagau]{matz_2023_b}
Matz,~F.; Nijssen,~J.; Jagau,~T.-C. Ab Initio Investigation of the Auger
  Spectra of Methane, Ethane, Ethylene, and Acetylene. \emph{J. Phys. Chem. A}
  \textbf{2023}, \emph{127}, 6147--6158\relax
\mciteBstWouldAddEndPuncttrue
\mciteSetBstMidEndSepPunct{\mcitedefaultmidpunct}
{\mcitedefaultendpunct}{\mcitedefaultseppunct}\relax
\EndOfBibitem
\bibitem[Jayadev \latin{et~al.}(2023)Jayadev, Ferino-Pérez, Matz, Krylov, and
  Jagau]{jayadev_2023}
Jayadev,~N.~K.; Ferino-Pérez,~A.; Matz,~F.; Krylov,~A.~I.; Jagau,~T.-C. The
  Auger spectrum of benzene. \emph{J. Chem. Phys.} \textbf{2023}, \emph{158},
  064109\relax
\mciteBstWouldAddEndPuncttrue
\mciteSetBstMidEndSepPunct{\mcitedefaultmidpunct}
{\mcitedefaultendpunct}{\mcitedefaultseppunct}\relax
\EndOfBibitem
\bibitem[Parravicini and Jagau(2023)Parravicini, and Jagau]{parravicini_2023}
Parravicini,~V.; Jagau,~T.-C. Interatomic and intermolecular Coulombic decay
  rates from equation-of-motion coupled-cluster theory with complex basis
  functions. \emph{J. Chem. Phys.} \textbf{2023}, \emph{159}, 094112\relax
\mciteBstWouldAddEndPuncttrue
\mciteSetBstMidEndSepPunct{\mcitedefaultmidpunct}
{\mcitedefaultendpunct}{\mcitedefaultseppunct}\relax
\EndOfBibitem
\bibitem[Jagau(2016)]{jagau_2016}
Jagau,~T.-C. Investigating tunnel and above-barrier ionization using
  complex-scaled coupled-cluster theory. \emph{J. Chem. Phys.} \textbf{2016},
  \emph{145}, 204115\relax
\mciteBstWouldAddEndPuncttrue
\mciteSetBstMidEndSepPunct{\mcitedefaultmidpunct}
{\mcitedefaultendpunct}{\mcitedefaultseppunct}\relax
\EndOfBibitem
\bibitem[Jagau(2018)]{jagau_2018}
Jagau,~T.-C. {Coupled-cluster treatment of molecular strong-field ionization}.
  \emph{J. Chem. Phys.} \textbf{2018}, \emph{148}, 204102\relax
\mciteBstWouldAddEndPuncttrue
\mciteSetBstMidEndSepPunct{\mcitedefaultmidpunct}
{\mcitedefaultendpunct}{\mcitedefaultseppunct}\relax
\EndOfBibitem
\bibitem[Heinrich-Josties \latin{et~al.}(2014)Heinrich-Josties, Pabst, and
  Santra]{Josties_2014}
Heinrich-Josties,~E.; Pabst,~S.; Santra,~R. Controlling the 2$p$ hole alignment
  in neon via the 2$s$-3$p$ Fano resonance. \emph{Phys. Rev. A} \textbf{2014},
  \emph{89}, 043415\relax
\mciteBstWouldAddEndPuncttrue
\mciteSetBstMidEndSepPunct{\mcitedefaultmidpunct}
{\mcitedefaultendpunct}{\mcitedefaultseppunct}\relax
\EndOfBibitem
\bibitem[Ueda \latin{et~al.}(1994)Ueda, Ganz, Harth, Bussert, Baier, Weber,
  Ruf, and Hotop]{Klar_1994}
Ueda,~D. K.~K.; Ganz,~J.; Harth,~K.; Bussert,~W.; Baier,~S.; Weber,~J.~M.;
  Ruf,~M.~W.; Hotop,~H. High-resolution measurement and quantum-defect analysis
  for the Ne nd' J= 1, 2 and 3 autoionizing resonances. \emph{J. Phys. B: At.
  Mol. Opt. Phys.} \textbf{1994}, \emph{27}, 4897\relax
\mciteBstWouldAddEndPuncttrue
\mciteSetBstMidEndSepPunct{\mcitedefaultmidpunct}
{\mcitedefaultendpunct}{\mcitedefaultseppunct}\relax
\EndOfBibitem
\bibitem[Min \latin{et~al.}(2008)Min, Lin-Fan, Cun-Ding, and Ke-Zun]{Min_2008}
Min,~G.; Lin-Fan,~Z.; Cun-Ding,~L.; Ke-Zun,~X. Optically Forbidden Excitations
  of 2s Electron of Neon Studied by Fast Electron Impact. \emph{Chin. Phys.
  Lett.} \textbf{2008}, \emph{25}, 3646\relax
\mciteBstWouldAddEndPuncttrue
\mciteSetBstMidEndSepPunct{\mcitedefaultmidpunct}
{\mcitedefaultendpunct}{\mcitedefaultseppunct}\relax
\EndOfBibitem
\bibitem[Chung \latin{et~al.}(1985)Chung, Lin, and Lee]{chung_1985}
Chung,~S.; Lin,~C.~C.; Lee,~E. T.~P. Rydberg states of the nitrogen molecule.
  \emph{J. Chem. Phys.} \textbf{1985}, \emph{82}, 342--352\relax
\mciteBstWouldAddEndPuncttrue
\mciteSetBstMidEndSepPunct{\mcitedefaultmidpunct}
{\mcitedefaultendpunct}{\mcitedefaultseppunct}\relax
\EndOfBibitem
\bibitem[Huber \latin{et~al.}(1993)Huber, Stark, and Ito]{Huber_1993}
Huber,~K.~P.; Stark,~G.; Ito,~K. Rotational structure in the Hopfield series of
  N$_2$. \emph{J. Chem. Phys.} \textbf{1993}, \emph{98}, 4471--4477\relax
\mciteBstWouldAddEndPuncttrue
\mciteSetBstMidEndSepPunct{\mcitedefaultmidpunct}
{\mcitedefaultendpunct}{\mcitedefaultseppunct}\relax
\EndOfBibitem
\bibitem[Woodruff \latin{et~al.}(1977)Woodruff, Marr, and
  Series]{woodruff_1977}
Woodruff,~P.~R.; Marr,~G.~V.; Series,~G.~W. The photoelectron spectrum of N$_2$
  and partial cross sections as a function of photon energy from 16 to 40 eV.
  \emph{Proc. Math. Phys. Eng. Sci.} \textbf{1977}, \emph{358}, 87--103\relax
\mciteBstWouldAddEndPuncttrue
\mciteSetBstMidEndSepPunct{\mcitedefaultmidpunct}
{\mcitedefaultendpunct}{\mcitedefaultseppunct}\relax
\EndOfBibitem
\bibitem[Warrick \latin{et~al.}(2016)Warrick, Cao, Neumark, and
  Leone]{Warrick_2016}
Warrick,~E.~R.; Cao,~W.; Neumark,~D.~M.; Leone,~S.~R. Probing the Dynamics of
  Rydberg and Valence States of Molecular Nitrogen with Attosecond Transient
  Absorption Spectroscopy. \emph{J. Phys. Chem. A} \textbf{2016}, \emph{120},
  3165--3174\relax
\mciteBstWouldAddEndPuncttrue
\mciteSetBstMidEndSepPunct{\mcitedefaultmidpunct}
{\mcitedefaultendpunct}{\mcitedefaultseppunct}\relax
\EndOfBibitem
\bibitem[Klinker \latin{et~al.}(2018)Klinker, Marante, Argenti,
  González-Vázquez, and Martín]{Klinker_2018}
Klinker,~M.; Marante,~C.; Argenti,~L.; González-Vázquez,~J.; Martín,~F.
  Electron Correlation in the Ionization Continuum of Molecules:
  Photoionization of N$_2$ in the Vicinity of the Hopfield Series of
  Autoionizing States. \emph{J. Phys. Chem. Lett.} \textbf{2018}, \emph{9},
  756--762\relax
\mciteBstWouldAddEndPuncttrue
\mciteSetBstMidEndSepPunct{\mcitedefaultmidpunct}
{\mcitedefaultendpunct}{\mcitedefaultseppunct}\relax
\EndOfBibitem
\bibitem[Little and Tennyson(2013)Little, and Tennyson]{Little_2013}
Little,~D.~A.; Tennyson,~J. An ab initio study of singlet and triplet Rydberg
  states of N$_2$. \emph{J. Phys. B: At. Mol. Opt. Phys.} \textbf{2013},
  \emph{46}, 145102\relax
\mciteBstWouldAddEndPuncttrue
\mciteSetBstMidEndSepPunct{\mcitedefaultmidpunct}
{\mcitedefaultendpunct}{\mcitedefaultseppunct}\relax
\EndOfBibitem
\bibitem[Woon and Dunning(1995)Woon, and Dunning]{woon_1995}
Woon,~D.~E.; Dunning,~J.,~Thom~H. {Gaussian basis sets for use in correlated
  molecular calculations. V. Core‐valence basis sets for boron through neon}.
  \emph{J. Chem. Phys.} \textbf{1995}, \emph{103}, 4572--4585\relax
\mciteBstWouldAddEndPuncttrue
\mciteSetBstMidEndSepPunct{\mcitedefaultmidpunct}
{\mcitedefaultendpunct}{\mcitedefaultseppunct}\relax
\EndOfBibitem
\bibitem[Kendall \latin{et~al.}(1992)Kendall, Dunning, and
  Harrison]{kendall_1992}
Kendall,~R.~A.; Dunning,~J.,~Thom~H.; Harrison,~R.~J. Electron affinities of
  the first‐row atoms revisited. Systematic basis sets and wave functions.
  \emph{J. Chem. Phys.} \textbf{1992}, \emph{96}, 6796--6806\relax
\mciteBstWouldAddEndPuncttrue
\mciteSetBstMidEndSepPunct{\mcitedefaultmidpunct}
{\mcitedefaultendpunct}{\mcitedefaultseppunct}\relax
\EndOfBibitem
\bibitem[Dunning(1989)]{dunning_1989}
Dunning,~J.,~Thom~H. Gaussian basis sets for use in correlated molecular
  calculations. I. The atoms boron through neon and hydrogen. \emph{J. Chem.
  Phys.} \textbf{1989}, \emph{90}, 1007--1023\relax
\mciteBstWouldAddEndPuncttrue
\mciteSetBstMidEndSepPunct{\mcitedefaultmidpunct}
{\mcitedefaultendpunct}{\mcitedefaultseppunct}\relax
\EndOfBibitem
\bibitem[Kaufmann \latin{et~al.}(1989)Kaufmann, Baumeister, and
  Jungen]{Kaufmann_1989}
Kaufmann,~K.; Baumeister,~W.; Jungen,~M. Universal Gaussian basis sets for an
  optimum representation of Rydberg and continuum wavefunctions. \emph{J. Phys.
  B: At. Mol. Opt. Phys.} \textbf{1989}, \emph{22}, 2223\relax
\mciteBstWouldAddEndPuncttrue
\mciteSetBstMidEndSepPunct{\mcitedefaultmidpunct}
{\mcitedefaultendpunct}{\mcitedefaultseppunct}\relax
\EndOfBibitem
\bibitem[Epifanovsky \latin{et~al.}(2021)Epifanovsky, Gilbert, Feng, Lee, Mao,
  Mardirossian, Pokhilko, White, Coons, Dempwolff, Gan, Hait, Horn, Jacobson,
  Kaliman, Kussmann, Lange, Lao, Levine, Liu, McKenzie, Morrison, Nanda,
  Plasser, Rehn, Vidal, You, Zhu, Alam, Albrecht, Aldossary, Alguire, Andersen,
  Athavale, Barton, Begam, Behn, Bellonzi, Bernard, Berquist, Burton, Carreras,
  Carter-Fenk, Chakraborty, Chien, Closser, Cofer-Shabica, Dasgupta,
  de~Wergifosse, Deng, Diedenhofen, Do, Ehlert, Fang, Fatehi, Feng, Friedhoff,
  Gayvert, Ge, Gidofalvi, Goldey, Gomes, González-Espinoza, Gulania, Gunina,
  Hanson-Heine, Harbach, Hauser, Herbst, Hernández~Vera, Hodecker, Holden,
  Houck, Huang, Hui, Huynh, Ivanov, Jász, Ji, Jiang, Kaduk, Kähler,
  Khistyaev, Kim, Kis, Klunzinger, Koczor-Benda, Koh, Kosenkov, Koulias,
  Kowalczyk, Krauter, Kue, Kunitsa, Kus, Ladjánszki, Landau, Lawler,
  Lefrancois, Lehtola, Li, Li, Liang, Liebenthal, Lin, Lin, Liu, Liu,
  Loipersberger, Luenser, Manjanath, Manohar, Mansoor, Manzer, Mao, Marenich,
  Markovich, Mason, Maurer, McLaughlin, Menger, Mewes, Mewes, Morgante,
  Mullinax, Oosterbaan, Paran, Paul, Paul, Pavošević, Pei, Prager, Proynov,
  Rák, Ramos-Cordoba, Rana, Rask, Rettig, Richard, Rob, Rossomme, Scheele,
  Scheurer, Schneider, Sergueev, Sharada, Skomorowski, Small, Stein, Su,
  Sundstrom, Tao, Thirman, Tornai, Tsuchimochi, Tubman, Veccham, Vydrov,
  Wenzel, Witte, Yamada, Yao, Yeganeh, Yost, Zech, Zhang, Zhang, Zhang, Zuev,
  Aspuru-Guzik, Bell, Besley, Bravaya, Brooks, Casanova, Chai, Coriani, Cramer,
  Cserey, DePrince, DiStasio, Dreuw, Dunietz, Furlani, Goddard,
  Hammes-Schiffer, Head-Gordon, Hehre, Hsu, Jagau, Jung, Klamt, Kong,
  Lambrecht, Liang, Mayhall, McCurdy, Neaton, Ochsenfeld, Parkhill, Peverati,
  Rassolov, Shao, Slipchenko, Stauch, Steele, Subotnik, Thom, Tkatchenko,
  Truhlar, Van~Voorhis, Wesolowski, Whaley, Woodcock, Zimmerman, Faraji, Gill,
  Head-Gordon, Herbert, and Krylov]{qchem}
Epifanovsky,~E. \latin{et~al.}  Software for the frontiers of quantum
  chemistry: An overview of developments in the Q-Chem 5 package. \emph{J.
  Chem. Phys.} \textbf{2021}, \emph{155}, 084801\relax
\mciteBstWouldAddEndPuncttrue
\mciteSetBstMidEndSepPunct{\mcitedefaultmidpunct}
{\mcitedefaultendpunct}{\mcitedefaultseppunct}\relax
\EndOfBibitem
\bibitem[Kaufman and Minnhagen(1972)Kaufman, and Minnhagen]{kaufman_1972}
Kaufman,~V.; Minnhagen,~L. Accurate Ground-Term Combinations in Ne i. \emph{J.
  Opt. Soc. Am.} \textbf{1972}, \emph{62}, 92--95\relax
\mciteBstWouldAddEndPuncttrue
\mciteSetBstMidEndSepPunct{\mcitedefaultmidpunct}
{\mcitedefaultendpunct}{\mcitedefaultseppunct}\relax
\EndOfBibitem
\bibitem[Hopfield(1930)]{hopfield_1930}
Hopfield,~J.~J. Absorption and Emission Spectra in the Region
  $\ensuremath{\lambda}600\ensuremath{-}1100$. \emph{Phys. Rev.} \textbf{1930},
  \emph{35}, 1133--1134\relax
\mciteBstWouldAddEndPuncttrue
\mciteSetBstMidEndSepPunct{\mcitedefaultmidpunct}
{\mcitedefaultendpunct}{\mcitedefaultseppunct}\relax
\EndOfBibitem
\bibitem[Ogawa and Tanaka(1962)Ogawa, and Tanaka]{ogawa_1962}
Ogawa,~M.; Tanaka,~Y. Rydberg absorption series of N$_2$. \emph{Can. J. Phys}
  \textbf{1962}, \emph{40}, 1593--1607\relax
\mciteBstWouldAddEndPuncttrue
\mciteSetBstMidEndSepPunct{\mcitedefaultmidpunct}
{\mcitedefaultendpunct}{\mcitedefaultseppunct}\relax
\EndOfBibitem
\bibitem[Krylov(2020)]{krylov_2020}
Krylov,~A.~I. From orbitals to observables and back. \emph{J. Chem. Phys.}
  \textbf{2020}, \emph{153}, 080901\relax
\mciteBstWouldAddEndPuncttrue
\mciteSetBstMidEndSepPunct{\mcitedefaultmidpunct}
{\mcitedefaultendpunct}{\mcitedefaultseppunct}\relax
\EndOfBibitem
\bibitem[Jayadev \latin{et~al.}(2023)Jayadev, Skomorowski, and
  Krylov]{jayadev_2023_b}
Jayadev,~N.~K.; Skomorowski,~W.; Krylov,~A.~I. Molecular-orbital framework of
  two-electron processes: Application to Auger and intermolecular Coulomb
  decay. \emph{J. Phys. Chem. Lett.} \textbf{2023}, \emph{14}, 8612--8619\relax
\mciteBstWouldAddEndPuncttrue
\mciteSetBstMidEndSepPunct{\mcitedefaultmidpunct}
{\mcitedefaultendpunct}{\mcitedefaultseppunct}\relax
\EndOfBibitem
\end{mcitethebibliography}

\end{document}